# The role of self-heating and hot-phonons in metallic single walled carbon nanotubes

Sayed Hasan[a], Ashraf Alam, Mark Lundstrom

*School of Electrical and Computer Engineering, Purdue University, West Lafayette,*

*Indiana 47907-1285*

## Abstract

The role of self-heating and hot-phonons on electronic transport in metallic single walled carbon nanotube is examined by solving a coupled electron-optical phonon (OP) Boltzmann transport equation and the classical heat equation. We find that self-heating reduces the high bias conductance, thus explaining the short OP mean free paths observed experimentally, while the non-equilibrium OP population reduces the overall current level. Comparison with experimental data shows that for tubes on a substrate, no significant population of hot phonons builds up under the normal range of applied bias. These results shed light on OP phonon decay paths in carbon nanotubes.

The electron-phonon interaction in metallic single walled carbon nanotubes (mSWNTs) has received considerable attention in recent years. High-field transport experiments[1,2,3] for mSWNTs on substrate suggest an optical phonon (OP) mean-free-path (mfp) of 10-15nm, while theoretically calculated values[3,4,5,6,7] lie in the range of 50-180nm. In this letter, we solve the electron Boltzmann transport equation (BTE), the OP BTE, and the classical heat equation self-consistently to explain the experimentally-observed short OP mfps in-terms of self-heating of the tube. Recent work by Pop et al.[8], and Kuroda et al.[9] also examined electro-thermal transport of mSWNTs, but the short OP mfps have not yet been explained. In this work, we describe for the first time coupled electron-OP transport in terms BTEs, and relax the local equilibrium between the OP and AP systems. The results shed new light on the OP lifetime in tubes on a substrate and tubes suspended above the substrate.

[a] Electronic address: hasanm@purdue.edu

We begin with a brief review of the physics of the electron-phonon interaction in metallic tubes, where conduction and valence bands cross at two Fermi points, and are well approximated by[1] $E_F = \pm \hbar \upsilon_F k$. As shown in Fig. 1a, four types of OP scattering[6] are possible in these bands, two forward (fs), and two backscattering (bs) processes. Assuming an armchair tube and following Mahan's approach,[5] the electron phonon coupling (EPC) parameter, $\Xi$, is calculated for each of these processes. Our approach differs from that of Mahan's, however, because instead of assuming a pure polarization for each phonon mode, the phonon eigenvectors are calculated using the zone folding[10] (ZF) method and then projected along the C-C bonds to calculate the EPCs, giving: $\Xi_\Gamma^{fs} \sim 9\sqrt{2}$ eV/A, $\Xi_\Gamma^{bs} \sim 9\sqrt{2}$ eV/A, $\Xi_K^{bs} \sim 18$ eV/A, and $\Xi_K^{fs} \sim 0$. Since phonon modes are not pure at the zone-boundary, the use of ZF is essential to calculate the EPC for the K-processes. Also, the normalization of eigenvectors increases EPC by a factor of $\sqrt{2}$ at the zone center as compared to Mahan's calculation. The phonon energy associated with the Γ-phonon is ~0.195eV, and for the K-phonon it is ~0.18eV. Assuming $d$ as tube diameter, the OP mfps[11] are next calculated as: $\ell_{\Gamma\text{-bs}} \sim 60d$, $\ell_{\Gamma\text{-fs}} \sim 60d$, and $\ell_{K\text{-bs}} \sim 28d$. Although the K-bs is the dominant scattering mode with 0.18eV energy, effects of other modes are also present to some extent, so it is more appropriate to compare experiments with an effective mfp calculated using Methissen's rule,

$$\ell_{eff}^{-1} = \ell_{K\text{-bs}}^{-1} + \left(\ell_{\Gamma\text{-bs}} + \ell_{\Gamma\text{-fs}}\right)^{-1} = 22d . \tag{1.1}$$

In our numerical simulations, however only a single effective K-bs mode of energy 0.18eV is considered for simplicity with mfp, $\ell$ used as an input parameter.

Three coupled transport equations are solved:

$$\partial_t f_\alpha = \pm\left\{-\upsilon_F \times \partial_x f_\alpha + \left(q\mathcal{E}/\hbar\right) \times \partial_k f_\alpha\right\} + S_\alpha^{in}(n,T) . \tag{1.2}$$

$$\partial_t n = G(q) - \upsilon_g \times \partial_x n - \left\{n(q) - n_{eq}(q,T)\right\}/\tau . \tag{1.3}$$

$$A \times \partial_x \left(\kappa(T)\partial_x T\right) - g \times (T - T_0) = \beta VI/L . \tag{1.4}$$

The first one is the electron BTE in 4 linear bands; the second one is the OP-BTE in the relaxation time approximation (RTA), and the third equation describes classical heat flow through the tube. In these equations, $\upsilon_F$, $\mathcal{E}$, $q$, $S_\alpha^{in}$, $G(q)$, $\upsilon_g$, $\tau$, $\kappa$, $g$, $A$ are respectively the Fermi velocity, electric field, magnitude of electron charge, the collision integral for net in-scattering, the OP generation rate, group velocity, and lifetime, the thermal conductivity of the tube, the thermal conductance of tube-substrate interface, and the heat conduction cross section. The OP relaxation time, $\tau$, represents OP to AP breakdown as well as the decay of OP to the substrate. Thus some fraction of the OP energy directly goes to the substrate, and the rest goes to the AP system. At steady state, the output power of the OP system is equal to the Joule heating, *VI*, and $\beta$ in Eq. (1.4) represents the fraction of it that goes to the AP system. In this work, β =1 is assumed for simplicity. Solution of Eq. (1.4) gives the temperature profile along the tube, which for longer tubes can be approximated reasonably well by an average temperature, as is evident from the profiles of Fig. 1c. For long tubes ($L >> \ell_{eff}$) with an average temperature and small OP group velocity, the electron and OP distribution function becomes spatially homogeneous, and $\partial_x$ terms in Eq. (1.2) and (1.3) can be safely ignored. If $\partial_x$ is ignored in Eq. (1.2), then an additional elastic scattering of mfp *L* is added to include the intrinsic contact resistance ($4q^2/h$). Thus, for an average *T* obtained by solving Eq. (1.4), Eqs. (1.2), and (1.3) are solved iteratively to calculate the Joule heating, *VI*. Using *VI* as an input, Eq. (1.4) is solved again to obtain new temperature, and the whole process is repeated until self-consistency is achieved. Figure 1b shows the computed electron distribution function for right moving and left moving electrons at ~4V bias. As the bias increases, the two electronic streams separate; at high bias, due to strong OP scattering, this separation becomes pinned at the phonon emission energy, giving current saturation in metallic tubes.

Fig. 2a shows the I-V characteristics for three different cases: i) the isothermal case, where only Eq. (1.2) is solved assuming T=300K ii) the self-heating only case where Eqs. (1.2) and (1.4) are solved, and $n(q)$ is determined using the Bose-Einstein distribution at the tube temperature, and iii) the hot-phonon case without the self-heating; here Eqs. (1.2), and (1.3) are solved with $\tau = 0.1 ps$, and T=300K. This particular value of $\tau$, is used only for illustrative purposes. Along with these three cases, the I-V characteristic with only the elastic scattering is also shown. Consider case i) first; here the I-V shows the characteristics features of OP emission, which is a) the low to moderate bias ($\hbar\omega L/qV \gg \lambda_{op}$) resistance, $R = V/I \approx R_0 \times (L/\lambda_{el} + qVL/\hbar\omega)$ is proportional to the bias[1], and b) the high-bias conductance is proportional to OP mfp. Since, the resistance is inversely proportional to phonon emission energy, $\hbar\omega$, the overall current level scales up proportionately with phonon emission energy. Now for case iii), where hot-phonon effect is present, the overall current level is smaller than case i) due to additional OP absorption scattering, and the effect is equivalent to having a lower OP emission energy. This is because, if there were no absorption the electrons would have traversed longer before it could emit an OP, but now that there is OP available, the electrons can gain energy by absorbing an OP and thereby cutting short the effective mfp, which is equivalent to having a smaller emission energy with the same mfp. It is interesting to note that much severe conductance reduction occurs at very low bias in case iii) due to the re-absorption of OPs emitted by the Fermi tail of the distribution function, which is in contrast to the experimental observation[1,2,3]. For case ii), on the other hand, where self-heating is present, the I-V does not change much at low-bias (because the temperature rise is small). At high bias, however, the temperature rise becomes significant, and the OP population increases at a much faster rate reducing the high-bias conductance compared to case i). This reduced conductance at high bias would be interpreted as a reduced OP mfp if self-heating effects were neglected. Figs. 2b, and 2c show the phonon distribution function and the variation of peak phonon population with applied bias respectively for cases ii) and iii).

Let us now compare experimental data obtained from three different sources [1,2,12] with the simulation with self-heating and hot-phonon effects turned on. Here $s$, $\ell$, and $\tau$ are used as input parameters. Fig. 3 shows the comparison between theory and experiment. To reproduce the data, $\ell_{op}$ = 20, 35, and 50nm are used for $d$ = 1, 2, and 2.4nm tubes respectively (The direct calculation of mfp using Eq. (1.1) gives mfps of 22, 44, and 52nm for these diameters). The important point, however, is that the data could only be matched by assuming an OP lifetime of ~5fs to effectively kill the off-equilibrium OP population generated by electron-phonon scattering. We conclude that for nanotubes on a substrate, the off-equilibrium OP population is negligibly small. The inset of Fig. 3 shows the simulated I-V using the same parameters, but with self-heating and hot-phonon effects turned off. For a $SiO_2$ substrate the simulated curved without self-heating is much higher than the data, indicating that significant self-heating is present in the data. For a $Si_3N_4$ substrate, however, simulation without self-heating reproduces the data, indicating little/no self heating in the data. Indeed, the thermal conductivity of $Si_3N_4$ is ~20 times larger than $SiO_2$ making the parameter, $g$ ( $g \propto k_{sub}/d$ [13] ) much larger for $Si_3N_4$, thus making it much harder to heat up.

The OP lifetime, which includes OP to (CNT) AP decay and the decay of OPs to the substrate due to substrate-SWNT coupling, is found to be extremely small, but it is consistent with previous experimental observations. Consistency with previous work can be established from the OP non-equilibrium parameter, $\alpha$ ( ratio between the OP to AP thermal resistance [12,14] ) in the following way

$$\alpha = \frac{R_{op}}{R_{ap}} \approx \frac{\tau/C_{op}}{1/gL} = \frac{gL}{C_{op}}\tau \rightarrow \tau = \frac{\alpha}{gL}C_{op}. \quad (1.5)$$

The OP branch specific heat is estimated as, $C_{op}(T) \sim Nk_B \left(x/e^{\hbar\omega/k_BT} - 1\right)^2$, where N is the number of excited phonon modes. Assuming a typical OP temperature of ~600K, the number of excited phonon modes becomes $N \sim k_B 600/2\pi\hbar\upsilon_F \sim 10^7$. Using the experimental data[12] we estimate, $\alpha = 0.01$ for tube on substrates. For a 1μm long tube, using these values with, $g = 0.2$W/m/K [8], $\tau$ becomes $\sim 10^{-15} s$. The effect of the substrate on the OP lifetime, which is likely to be the dominant factor here, has also been observed experimentally[14]. Note that mSWNTs suspended over a trench show significantly lower current[12], and to match these data, we need a significantly large OP lifetime of $\tau \sim 0.1$ps.

In conclusion, a self-consistent electro-thermal calculation was done to study the effect of self-heating and hot-phonons on the current-voltage characteristics of mSWNTs on a substrate. We found that self-heating is responsible for the observed low conductance at high bias but that the off-equilibrium OP population is negligibly small. With any significant population of off-equilibrium optical phonons, the shape of the measured IV characteristics cannot be explained. For nanotubes on a SiO$_2$ substrate, the I-V characteristics only bear the signature of self-heating, and for Si$_3$N$_4$ substrates, the available data[12] suggest little or no self-heating consistent with the higher thermal conductivity of Si$_3$N$_4$.

Hasan, Alam, LundstromThis work was supported by Intel Corporation with computational support provided by the NSF-funded Network for Computational Nanotechnology under grant no. EEC-0228390. The authors benefitted form extensive discussions with H. Dai and E. Pop from Stanford, and Dmitri Nikonov from Intel, and are thankful to S. Kumar at Purdue for suggesting the solution of heat equation.8

Hasan, Alam, Lundstrom

**List of Figure Captions**

Fig. 1. (a) Linear metallic bands in mSWNTs. Four scattering processes Γ-fs, Γ-bs, K-fs, and K-bs are also shown. (b) Electron distribution function, $f_R$, and $f_L$ for positive and negative moving electrons. (c) Temperature profile along the tube with 20μW Joule heating as input. Thermal conductivity is assumed to be as $3000 \times 300/T$ W/m/K.

Fig. 2. (a) I-V shown for three different cases. Case i) isothermal, case ii) with self-heating only, and case iii) non-equilibrium OP only with $\tau = 0.1 ps$. Also shown is the I-V with only elastic scattering. (b) OP distribution function for case ii) and iii) at 1.6V bias. (c) The peak OP population with bias for case ii) and iii).

Fig. 3. Comparison of the simulation with experimental data[2][1][12]. For 700nm tube, $d = $ 2nm, $\ell_{con} = \infty$, $\ell_{op} = 35nm$, $g = 0.1$ W/m/K. For 1μm tube, $d = $ 1nm, $\ell_{con} = 300nm$, $\ell_{op} = 20nm$, $g = 0.2$ W/m/K. These two are for SiO$_2$ substrate. Note that, for $d$=1nm tube, g is larger than it is for $d$=2nm tube, as expected [13]. The data for Si$_3$N$_4$ substrate is shown on the negative voltage axis. The parameters are: $\ell_{con} = 950nm$, $\ell_{op} = 50nm$, $g = 1.8$ W/m/K. Higher $g$ is used due to larger thermal conductivity of Si$_3$N$_4$. In all cases, $\ell_{ap} = 2\mu m$, $\hbar\omega = 0.18 eV$, and $\tau = $ 5fs are assumed. Here $\ell_{con}$ represents extrinsic contact resistance, and $1/\ell_{el} = 1/L + 1/\ell_{con} + 1/\ell_{ap}$. Inset shows the simulation assuming no self-heating, and hot-phonon effects.

## List of Figures

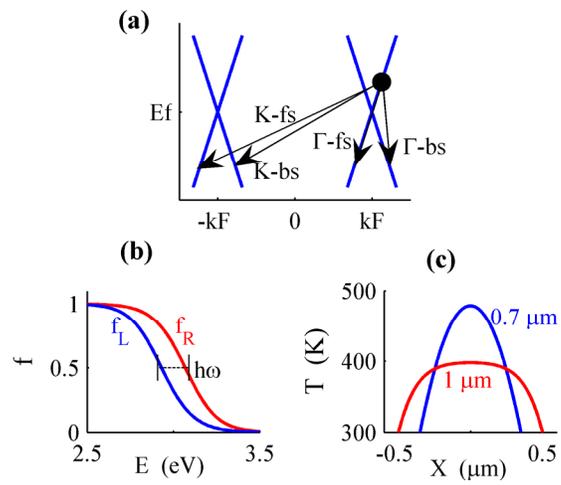

Figure 1

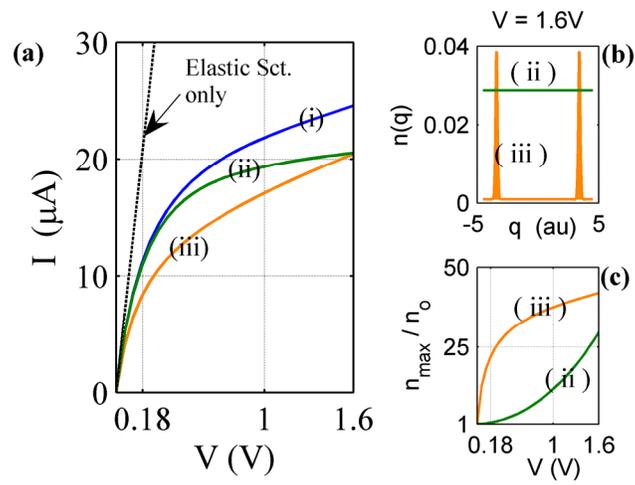

Figure 2

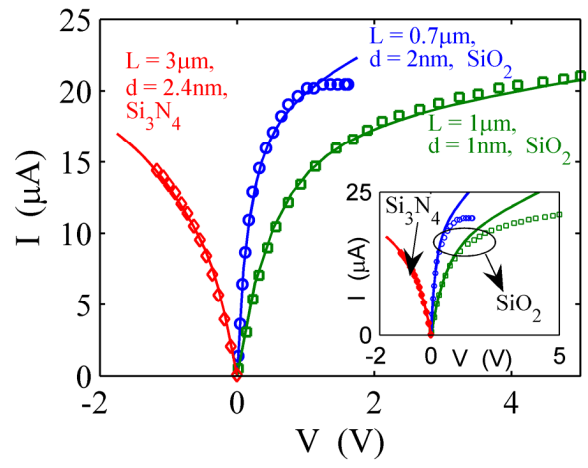

Figure 3